# Optical Microscopy of Soft Matter Systems

*Taewoo Lee[1], Bohdan Senyuk[1], Rahul P. Trivedi[1], and Ivan I. Smalyukh[1, 2,*]*

[1]Department of Physics and Liquid Crystal Materials Research Center, University of Colorado, Boulder, Colorado 80309, USA
[2]Renewable and Sustainable Energy Institute, National Renewable Energy Laboratory and University of Colorado, Boulder, Colorado 80309, USA
*Email: Ivan.Smalyukh@colorado.edu

## Table of Contents

## 10.1. Introduction

The fast-growing field of soft matter research requires increasingly sophisticated tools for experimental studies. One of the oldest and most widely used tools to study soft matter systems is optical microscopy. Recent advances in optical microscopy techniques have resulted in a vast body of new experimental results and discoveries. New imaging modalities, such as nonlinear optical microscopy techniques that were developed to achieve higher resolution, enable soft matter research at length scales ranging from the molecular to the macroscopic.

The aim of this chapter is to introduce a variety of optical microscopy techniques available to soft matter researchers, starting from basic principles and finishing with a discussion of the most advanced microscopy systems. We describe traditional imaging techniques, such as bright field and polarizing microscopy, along with state-of-the-art three-dimensional (3D) imaging techniques, such as fluorescence confocal and nonlinear optical microscopies. Different approaches are discussed along with their applications in the study of soft matter systems by providing typical examples.

The chapter is organized as follows. Section 10.2 deals with basic principles and parameters important to all optical microscopy modalities. Sections 10.3 through 10.5 describe traditional optical microscopy techniques. Fluorescence and confocal microscopies are discussed in sections 10.6 through 10.8. Section 10.9 provides an extended account of nonlinear optical microscopy techniques. An example of particle tracking technique that utilizes a pre-engineered point spread function and its application in imaging is detailed in section 10.10. Since many experiments in soft matter systems require simultaneous non-contact optical manipulation and 3D imaging, integration of these techniques is immensely useful and is discussed in section 10.11. We conclude our review on optical microscopy and discuss future prospectives of its use in the study of soft matter systems in section 10.12.

## 10.2. Basics of optical microscopy

The history of optical microscopy began centuries ago. It is difficult to say who invented the compound optical microscope, but it has been used extensively for research in different branches of science starting from the beginning of the 17th century. Since that time optical microscopy has developed into its own research field and industry at the forefront of science and engineering. The developments of new techniques in optical microscopy have often led to significant breakthroughs in many different scientific fields.

The simplest optical imaging system that can be used as a microscope consists of two converging lenses: an objective lens and an eyepiece. The optical train of this microscope and its corresponding ray diagram are shown in Fig. 1(a). Illumination light transmitted through the sample is collected by the objective lens and transferred to the eyepiece, forming an image on the retina of the observer's eyes. Various types of light detecting devices, such as charge-coupled device (CCD) cameras, photodiodes, avalanche photodiodes, photomultiplier tubes, and other optical sensors are nowadays widely used for collecting the image. Electronic scanning systems such as galvano-mirrors and acousto-optic deflectors or fast confocal illumination systems such as Nipkow discs are often utilized in modern optical microscopes and imaging systems.

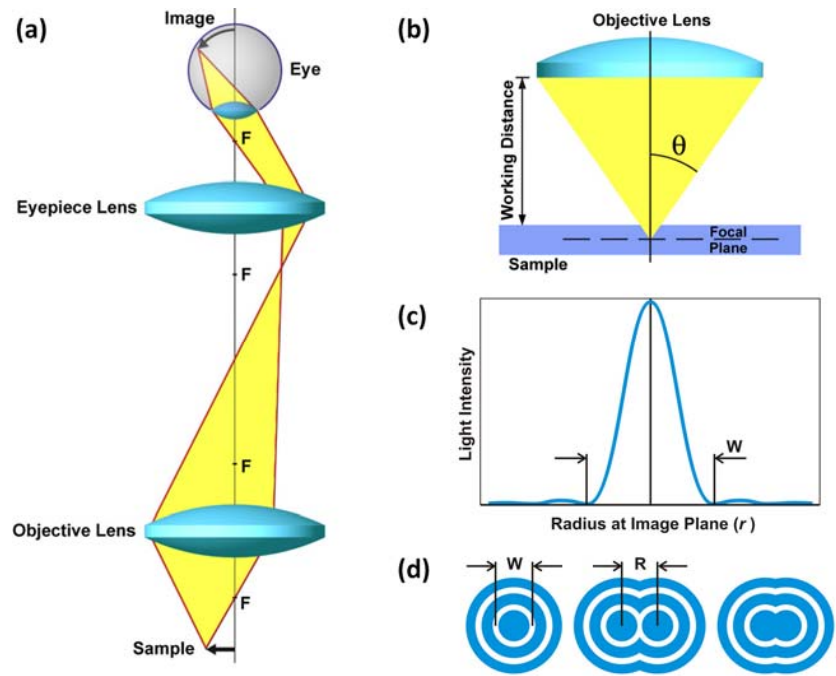

**Figure 1.** Principles of imaging with an optical microscope: (a) ray diagram of the simplest two-lens microscope; (b) definition of parameters of an objective lens; (c) point spread function with the first minimum at $r$=W/2; (d) Airy patterns in the image plane. "F" marks the foci of lenses. "R" shows the separation distance between the centers of Airy discs.

Objective lenses are perhaps the most essential optical components and are characterized by parameters like numerical aperture and working distance, which largely define their performance. The numerical aperture (NA) of the objective determines the range of angles over which the objective lens can accept light and is defined as

$$\mathrm{NA} = n \cdot \sin\theta \,,$$

where $n$ is the refractive index of the medium between the objective lens and the sample, and $\theta$ is the half-angle of the cone of light accessible for collection by the objective [Fig. 1(b)]. While NA is less than unity for air objectives, oil immersion objectives can have NA close to 1.5. Another important parameter characterizing the objective is its working distance (WD), defined as the distance from the front lens of the objective to the surface of the sample when the inspected area is in focus [Fig. 1(b)]. The depth up to which the sample can be imaged is often restricted by the available WD. Objectives with larger NA usually have smaller WD and vice versa.

Numerical aperture determines important microscope characteristics such as the point spread function (PSF) and, consequently, the resolution. The PSF [Fig. 1(c)] can be thought of as the light intensity distribution in the image acquired by the microscope from a point source, and is given by [1]

$$\text{PSF}(r) = \left[\frac{2J_1(ra)}{ra}\right]^2,$$

where $a = 2\pi \, \text{NA} / \lambda$, $J_1$ is the Bessel function of the first kind, $\lambda$ is the wavelength of light and $r$ is the distance from the center of the peak in light intensity [Fig. 1(c)]. In the image plane of a standard optical system, the PSF is shaped as the Airy diffraction pattern with the first minimum at $r = \text{W}/2$ [Fig. 1(d)]. Two Airy discs separated by the distance R $\geq$ W/2 [Fig. 1(d) can be resolved into separate entities, but not at smaller R [2]. This limit is often called the Rayleigh criterion or the diffraction limit and defines the lateral resolution of the objective as $r_{lateral} = 0.61\lambda / \text{NA}$, i.e., the resolution in the plane orthogonal to the microscope's optical axis. From the definition of $r_{lateral}$ it follows that objectives with higher NA can resolve finer details, however, even for high-NA objectives, the lateral resolution can be only slightly smaller than the wavelength of light used for imaging. Although conventional optical microscopes have poor resolution along the optical axis of a microscope, recently introduced optical imaging techniques, such as confocal and nonlinear optical microscopies, allow for optical imaging with high axial resolution. The axial resolution of an optical system will be introduced later when discussing fluorescence confocal microscopy in section 10.7. Modern imaging approaches allowing one to overcome the diffraction limit for both radial and axial directions will be discussed in section 10.12.

## 10.3. Bright field and dark field microscopy

Bright field imaging is perhaps the simplest of all optical microscopy methods. A sample is illuminated by unpolarized white light and the contrast in the image results from direct interaction of the probing light with the sample (absorption, refraction, scattering, reflection, etc.). Figure 2(a) shows a simple schematic diagram of transmission mode (dia-illumination) bright field microscopy, where the illumination light is focused onto the sample by a condenser lens with numerical aperture $NA_{cond}$ and the transmitted light is collected by the objective lens with numerical aperture $NA_{obj}$. High quality imaging can be performed at optimum illumination of the sample using the so-called "Köhler illumination" [3]. The ratio between the NAs of the objective and condenser lenses affects the resolution and contrast of imaging. The resolution in the microscope image is optimized when the NAs of condenser and objective are equal. Using objectives with $NA_{obj}$ higher than $NA_{cond}$ increases the image contrast but decreases resolution.

The example images obtained with bright field microscopy show the texture of toroidal focal conic domains in a thin film of a smectic A liquid crystal spread over a glycerol surface [Fig. 2(b)] and an array of *P. aeruginosa* cells trapped by an array of infrared (invisible in the image) laser beams [Fig. 2(c)]. The contrast in these images [Fig. 2(b,c)] results mostly from refraction and scattering of light at the boundaries of objects having refractive index different than that of the surrounding medium. One of the main limitations of bright field imaging is that samples with weak spatial variation of the refractive index or absorption produce images with poor contrast.

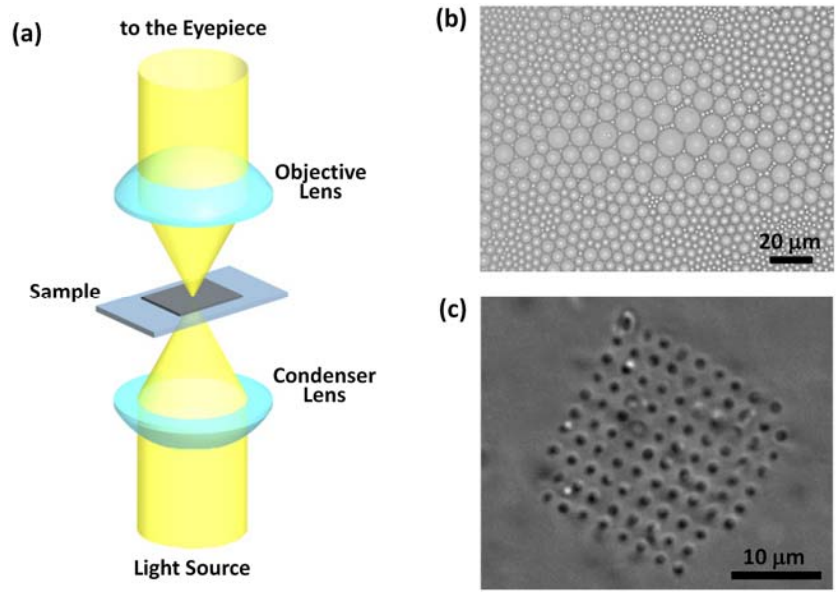

**Figure 2.** Bright field microscopy diagram (a) and textures: (b) toroidal focal conic domains in a smectic liquid crystal (8CB) film spread over a glycerol surface; (c) an array of *P. aeruginosa* cells trapped by an array of infrared laser beams.

Dark field microscopy is a method in which the image is formed by collecting only the light scattered by the sample. All unscattered light coming directly from the illumination source is excluded from the image, thus forming a dark background [Fig. 3(a)]. To achieve this, the specimen is illuminated with a hollow cone of light [Fig. 3(a)] formed by a special condenser with an opaque light-stop blocking the direct light. The oblique illumination light is scattered by objects in the field of view and only the scattered light is collected by the objective, while unscattered direct light is blocked. For optimized imaging in this optical arrangement, $NA_{cond}$ should be larger than $NA_{obj}$.

Dark field microscopy is typically used for imaging of samples with sparse scattering objects dispersed in a non-scattering medium, and is especially useful in the study of objects having size below the diffraction limit. For example, this technique is employed when investigating systems ranging from red blood cells to plasmonic metal nanoparticles [4]. Figure 3(b) shows dark field image of individual gold rod-like colloidal particles dispersed in a smectic A liquid crystal. The color and contrast in the image is formed due to absorption and scattering of the white illumination light by rod-shaped gold nanoparticles shown in the inset of Fig. 3(b).

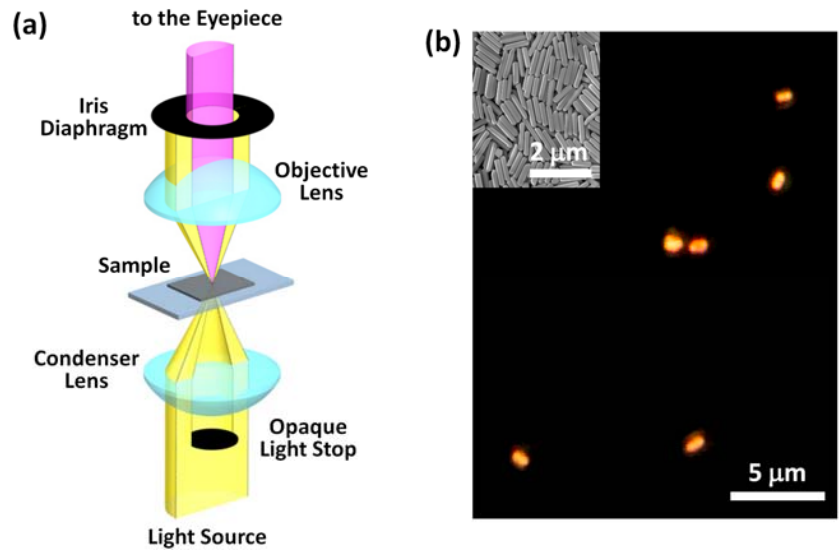

**Figure 3.** Dark field microscopy diagram (a) and texture of gold rod-like colloids (b) dispersed in a smectic (8CB) homeotropic liquid crystal cell. Inset in (b) shows the transmission electron microscopy image of gold rod-like colloids (image courtesy of Nanopartz Inc.).

## 10.4. Polarizing microscopy

Polarizing microscopy (PM) is an imaging method that uses polarized light and is usually utilized for imaging birefringent and optically active materials, such as liquid crystals, minerals and crystals [3]. Typically, the illumination is linearly polarized by a "polarizer" and upon transmitting through the sample, is passed to another polarizer (called "analyzer") [Fig. 4(a)]. The analyzer is usually kept parallel or orthogonal to the polarizer. The intensity of the light transmitted by the sample between two crossed polarizers depends on the angle $\beta$ between the polarization of light and the optic axis of birefringent sample and phase retardation $2\pi\Delta nd/\lambda$ introduced by a uniform sample of thickness $d$ [5]:

$$I = I_0 \sin^2(2\beta) \sin^2(\pi\Delta nd/\lambda),$$

where $I$ and $I_0$ is the intensity of transmitted and incident light, respectively, and $\Delta n$ is the birefringence. If the sample is viewed in the same position between parallel polarizers instead, the intensity of transmitted light becomes $I = I_0\left[1 - \sin^2(2\beta)\sin^2(\pi\Delta nd/\lambda)\right]$. In both cases, linearly polarized incident light is split by the birefringent sample into two components, extraordinary and ordinary, propagating within the sample at different speeds, which results in elliptical polarization of light upon exiting the sample [Fig. 4(a)]. The analyzer probes the polarization state of the transmitted light as altered by the birefringent medium. The intensity pattern in the image is then used to deduce the spatial variations of the orientation of the optic axis and/or value of birefringence in the lateral plane of an optically anisotropic sample.

An example image acquired with PM between crossed polarizers [Fig. 4(b)] shows the texture of a thin film of a nematic liquid crystal [6] spread over a liquid substrate such as glycerol. The contrast and color in the image result from the phase shift between the two propagating modes. The dark brushes in the image correspond to the locations where the local orientation of optic axis in the sample is along the polarizer or analyzer. Vivid colors are produced as a result of interference when the phase retardation is of the order of $2\pi$. The color pattern can be utilized to deduce the values of optical anisotropy, orientation of the optic axis, and the film thickness.

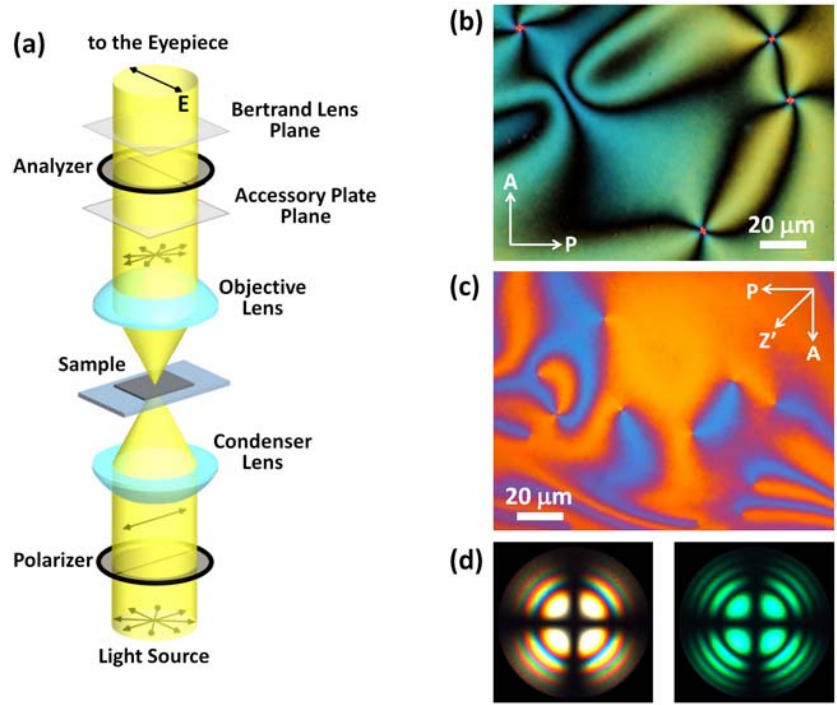

**Figure 4.** Polarizing microscopy: (a) diagram; (b) Schlieren texture observed in the thin film of nematic liquid crystal spread over glycerol surface; (c) Schlieren texture observed with the red plate inserted; (d) conoscopic images obtained for a homeotropic uniaxial smectic A (8CB) sample with a Bertrand lens inserted under the white (left) and green (right) light. Arrows P, A and Z' show a direction of polarizer, analyzer and "slow axis" of the red plate, respectively.

The capabilities of the PM technique in the study of birefringent samples can be greatly expanded by using accessory birefringent plates and wedges in the optical path. The orientation of the optic axis of a birefringent material can be determined by introducing a full wavelength (typically ~530 nm) retardation plate (the so-called "red plate") into the optical train [3, 5] after the sample and objective [Fig. 4(a)]. In the example shown in Fig. 4(c), the in-plane orientation of the local liquid crystal optic axis is parallel to the "slow axis" of the red plate in the bluish regions and perpendicular to it in the yellowish regions. The birefringence of the sample and its sign can be determined using a quartz wedge, quarter-wave retardation plate (Senarmont method), or the Berek compensator [3, 5]. The principles behind the use of all of these accessory plates are based on probing the addition or subtraction of the phase retardations due to the retardation plate or wedge (with known orientation of the optic axis and known phase retardation) and the studied sample. The analysis of the resulting phase retardation patterns yields information complementary to that obtained by imaging the sample between a pair of polarizers alone.

The introduction of a Bertrand lens into the optical train of the microscope enables an imaging mode known as "conoscopy" [Fig. 4(d)]. The Bertrand lens is inserted in the optical train after the analyzer [Fig. 4(a)] and brings the objective's back aperture plane into the focus. In this arrangement, the sample is illuminated by a strongly convergent cone of light, which provides an interference image [Fig. 4(d)] formed by light passing through the sample at different angles. Conoscopic arrangement of a polarizing microscope allows determination of the type of birefringent materials (i.e., uniaxial or biaxial), orientation of the optic axes, and the sign of birefringence [3, 5].

### 10.5. Differential interference contrast and phase contrast microscopy

Most common optical microscopy techniques used to enhance the contrast in transparent samples with weak spatial variation of the refractive index are differential interference contrast (DIC) and phase contrast microscopies. In the DIC technique, image contrast results from the gradient of the refractive index within the sample. A sample is placed between crossed polarizers and two beamsplitters, the Nomarski-modified Wollaston prisms [Fig. 5(a)], such that the prisms' optic axis is kept at $45^{\circ}$ with respect to the crossed polarizers. The polarized illumination light is separated by the first beamsplitter into two orthogonally polarized and spatially displaced (sheared) components [red and blue in Fig. 5(a)]. Each component propagates along a different path, giving rise to optical path difference upon exiting the sample. After being collected by an objective, the two components are recombined by the second Nomarski prism and transmitted through the analyzer to the detector. As a result, the contrast in DIC images depends on the path length gradient along the shear direction; this enhances the contrast due to edges or interfaces between sample areas with different refractive indices. Figure 5(b) shows a DIC image of an oil-in-water emulsion [7]. In this example, the interface between water and oil droplets is highlighted due to the difference in their refractive indices and the resultant gradient of optical paths along the shear direction. The image contrast increases as this gradient becomes steeper.

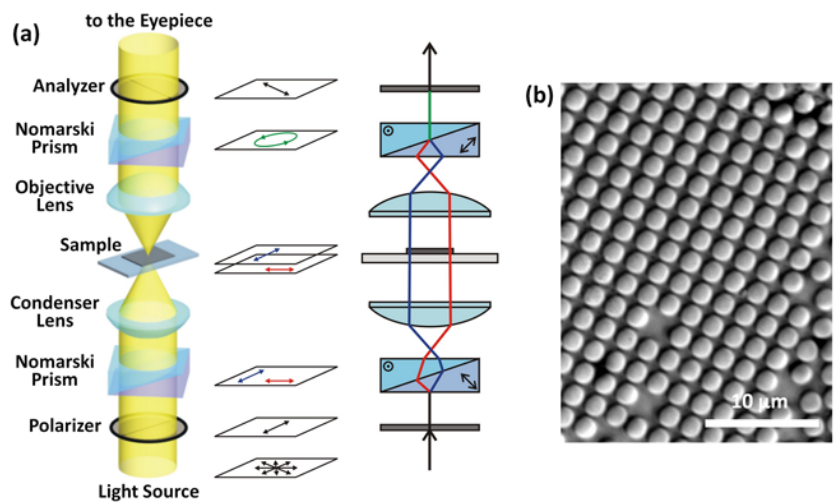

**Figure 5.** Differential interference contrast microscopy (a) diagram and (b) texture of an oil-in-water emulsion (image courtesy of V. N. Manoharan).

Phase contrast microscopy technique transforms small spatial variations in phase into corresponding changes in the intensity of transmitted light. The diagram of phase contrast microscope's optical train is presented in Fig. 6(a). The illumination light passes through an annular ring placed before the condenser lens and is focused on the sample. It either passes through undeviated [yellow in Fig. 6(a)] or is diffracted with changed phase [violet in Fig. 6(a)], depending on the composition of the sample. Both undeviated and diffracted light is collected by the objective and transmitted to the eyepiece through the phase ring, which introduces an additional phase shift to the undeviated reference light. Parts of the sample having different refractive indices appear darker or brighter compared to the uniform background, forming the phase contrast image. Phase contrast imaging is insensitive to polarization and birefringence effects. This is a major advantage when examining living cells and a number of other soft matter systems. Figure 6(b) shows the texture of elongated mycelium hyphae of a common *Aspergillus* mold confined within microfluidic channels [8] obtained with the phase contrast technique.

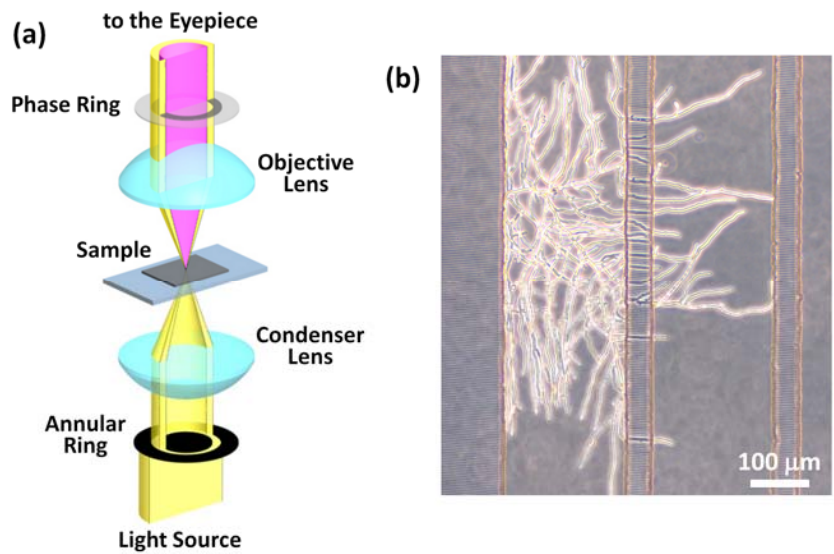

**Figure 6.** Phase contrast microscopy: (a) diagram and (b) the image of elongating mycelium hyphae of a common *Aspergillus* mold confined within polydimethylsiloxane-based microfluidic channels (image courtesy of L. Millet).

## 10.6. Fluorescence microscopy

The optical microscopy techniques described so far are based on transmission, absorption, refraction, or scattering of light. Fluorescence microscopy is a method based on the fluorescence phenomenon, where absorption of light by fluorescent molecules of dyes or fluorophores results in the emission of light at a longer wavelength [Fig. 7(a,b)]. The dye molecules absorb light of a specific wavelength [Fig. 7(b)], which results in a transition to a higher energy level known as the excited state [Fig. 7(a)]. After a short time delay (determined by the lifetime of molecules in the excited state, typically in the range of nanoseconds) the molecule returns to the initial ground state [Fig. 7(a)] with emission of light of wavelength longer than that of the absorbed light [Fig. 7(b)]. The difference between the wavelengths of maximum absorption and fluorescence (called the Stokes shift) is caused by loss of part of the absorbed energy due to non-radiative processes. This wavelength difference makes it possible to effectively separate the excitation and emission signals by use of optical filters.

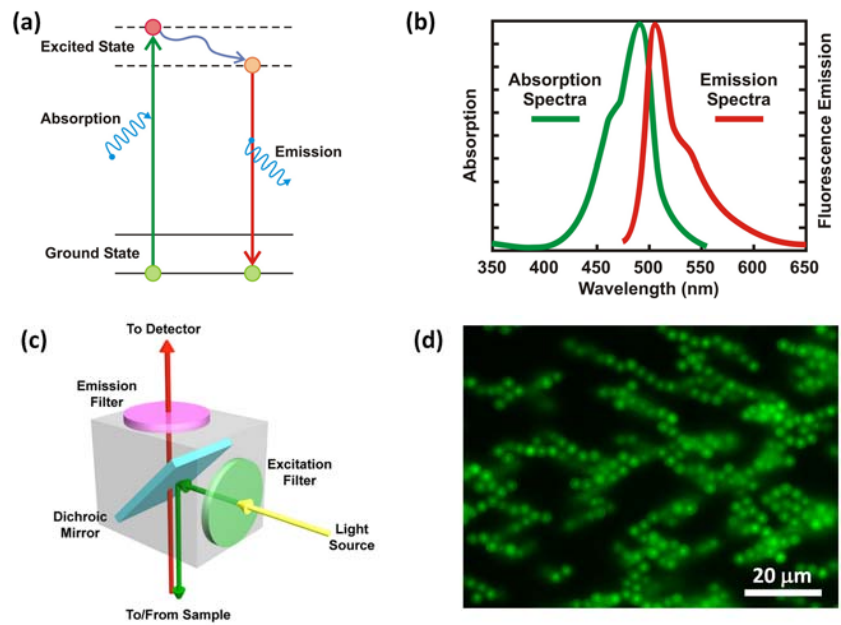

**Figure 7.** Fluorescence microscopy: (a) Jablonski diagram; (b) typical absorption and emission spectra of a fluorescent dye; (c) filters cube; (d) fluorescence texture of spherical microparticles tagged with a dye and dispersed in a nematic liquid crystal.

The basic function of a fluorescence microscope is to irradiate the sample with light of specific wavelength, and then to separate the fluorescence emission signal from the excitation light of much stronger intensity. In the fluorescence microscope setup, this is typically achieved with a filter cube consisting of excitation and emission filters and a dichroic mirror [Fig. 7(c)]. The constituents of interest in a sample are labeled with one or more fluorescent dyes. The excitation wavelength is selected by the excitation filter, reflected by the dichroic mirror to the sample, and is absorbed by the dye molecules in the sample. The fluorescent light emitted by the dye molecules transmits through the dichroic mirror and, after being separated from the excitation light by the emission filter [Fig. 7(c)], is collected by the photodetector. This forms an image with the bright areas corresponding to the sample regions marked with the fluorescent dye and the dark areas corresponding to the regions without the dye [Fig. 7(d)]. A fluorescence microscopy image in Fig. 7(d) shows colloidal structures formed in a nematic liquid crystal by melamine resin microspheres labeled with a fluorescent dye which emits green fluorescence light [9].

## 10.7. Fluorescence confocal microscopy

A more advanced technique, which offers 3D imaging capability by combining the features of fluorescence and confocal microscopies, is fluorescence confocal microscopy (FCM) shown in Fig. 8. The main feature of confocal microscopy is that the inspection region at a time is a small voxel (volume pixel) and the signal arising from the neighboring region is prevented from reaching the detector by having a pinhole in the detection image plane [Fig. 8(a)]. The inspected voxel and the pinhole are confocal, so that only the light coming from the probed voxel reaches the detector. This enables diffraction-limited imaging resolution not only in the lateral plane but also along the optical axis of the microscope. It is thus possible to construct a 3D image of the sample by scanning voxel by voxel within the volume of interest. In FCM, the contrast and sensitivity of confocal microscopy is greatly enhanced as the fluorescent dye used to tag the sample absorbs light at the wavelength of excitation laser beam and emits at a longer wavelength. This allows for imaging of fluorophore-tagged samples with diffraction-limited resolution in the axial plane of the microscope, a capability not provided by the conventional fluorescence microscopy.

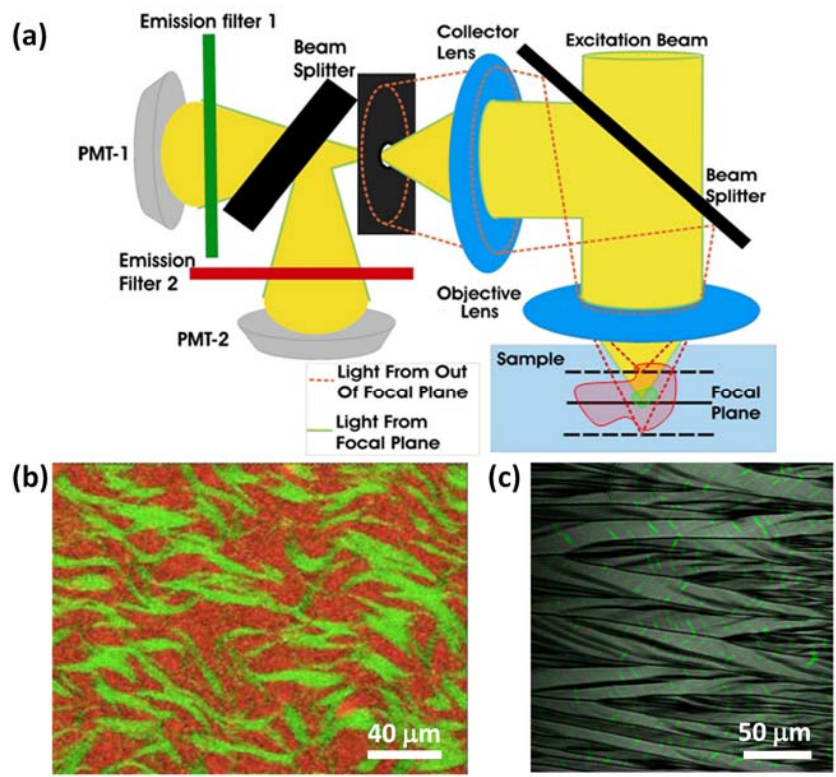

**Figure 8.** Fluorescence confocal microscopy: (a) principal diagram of a two-channel FCM setup; (b) image of phase-separated domains of F-actin (green) and DNA (red) labeled by two different dyes [11]; (c) polarizing microscope texture of the pattern formed in a drying droplet of aqueous DNA and colocalized with the fluorescence confocal signal from a small number of molecules marked by fluorescent dye (green) [12].

Figure 8(a) illustrates the basic principles of FCM. The combined features of fluorescence and confocal microscopies offer enhanced lateral resolution in isotropic samples, $r_{lateral} = 0.44\lambda$ / NA. The axial resolution of FCM is $\Delta z = 1.55 n\lambda$ / $NA^2$, where $n$ is the refractive index of the medium between the objective lens and the sample [10]. These theoretical limits of resolution, however, are rarely achieved in experiments because of different kinds of sample inhomogeneities, light scattering, spherical and chromatic aberrations, sample birefringence, index mismatches at dielectric interfaces along the light path. These and other effects can significantly worsen both axial and lateral resolution of FCM. For example, birefringent samples like liquid crystals cause the excitation beam to split into two separate beams which are focused at different spots, thus degrading the spatial resolution and making it dependent on the depth of imaging and sample birefringence.

Similar to conventional fluorescence microscopy (although we did not provide examples of this in the previous section), FCM enables multi-color imaging of sample composition patterns. Different constituents of the studied sample can be labeled with different dyes tailored to have different excitation and emission wavelengths and the resultant fluorescence patterns from each of them can be collocated to form separate images or overlaid to form a single multi-color image. For example, Figure 8(b) shows an image of phase-separated domains in a DNA/F-actin biopolymer mixture with the DNA and F-actin each labeled with a different dye [11]. Likewise, to obtain complementary information, one can use transmission mode bright field microscopy or polarizing microscopy, or a number of other optical imaging modalities to obtain images that can then be overlaid with the FCM images. This is demonstrated by an example of FCM textures [Fig. 8(c)] of a DNA pattern at the perimeter of dried drop overlaid with the PM image of the same sample area [12]. Multi-color imaging is especially useful in the study of composite soft matter systems.

Since imaging with a conventional confocal microscope is relatively slow, it is most often utilized in the study of stationary structures. To study dynamic processes in soft matter systems, researchers have recently started to utilize fast confocal microscopy systems, which can be implemented by use of the fast laser scanning with acousto-optic deflectors (instead of galvano-mirrors) or using the Nipkow disc. One example of implementation of the fast confocal imaging system, shown in Fig. 9, employs a rotating Nipkow disc having thousands of pinholes, supplemented by a coaxial disc with micro-lenses. The two discs are mechanically connected and are rotated together by an electrical motor. The sample is scanned by thousands of excitation beams at once, and the resultant speed of FCM imaging is higher by orders of magnitude as compared to that of a conventional confocal microscope. The vertical refocusing is performed by a fast piezo z-stepper drive that is capable of an accurate (typically ~ 50 nm or better) vertical position setting. The speed of imaging can reach 100-1000 frames per second, although it depends on many factors, such as the needed contrast (i.e., the integration time of the fluorescence signal), size of the scanned area, and the frame-rate of the used CCD camera. Thus,

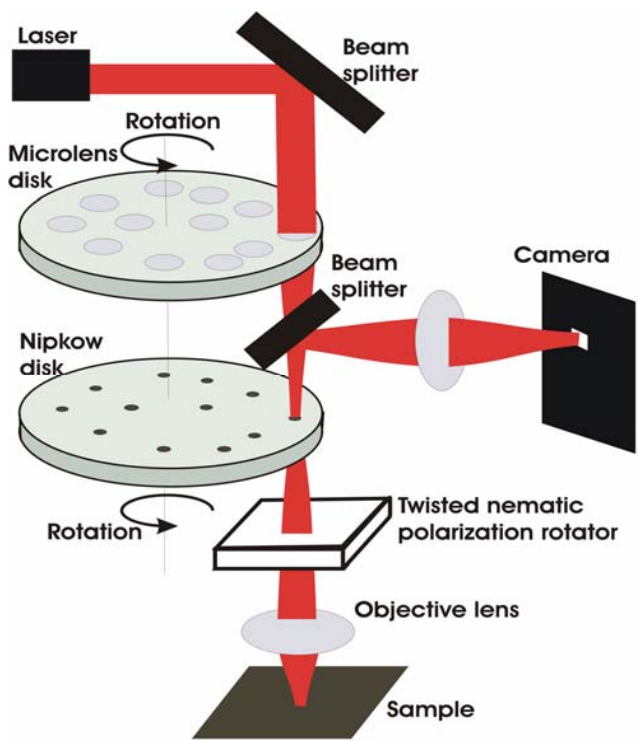

Nipkow disc based FCM allows one to decipher dynamics of colloidal structures and other soft matter systems with up to millisecond temporal resolution. For example, fast FCM has been widely used to simultaneously localize multiple (millions) dye-marked particles and has provided deep insights into the physics of colloidal aggregation and phase transition in colloidal fluids, gels, glasses, etc. [13].

**Figure 9.** A schematic of Nipkow disc fluorescence confocal microscopy.

**10.8. Fluorescence confocal polarizing microscopy**

The FCM technique described above visualizes the distribution of fluorescent dye, providing insight into the spatial distribution of different dye-tagged constituents throughout the sample. The orientational order of anisotropic soft matter systems can be probed by combining conventional FCM with the capability of polarized light excitation and detection. In fluorescence confocal polarizing microscopy (FCPM) [10], this is commonly achieved by annexing the FCM setup with a polarization rotator introduced before the objective lens (Fig. 10), which enables controlled polarized excitation. In addition, FCPM requires that the specimen be stained with anisometric dye molecules which, on average, align parallel or perpendicular to the molecules of the studied "host" material. FCPM signal strongly depends on the angle between the transition dipole moment of the dye molecules and the polarization of excitation light. The intensity of fluorescence is maximized when the linear polarization of excitation light is parallel to the transition dipoles of excitation and fluorescence of the dye molecules in the sample, and is at minimum when the polarization is perpendicular to the transition dipoles. This strong orientational dependence of the measured fluorescence signal allows one to decipher the 3D molecular orientation patterns and liquid crystal director fields directly from the FCPM images.

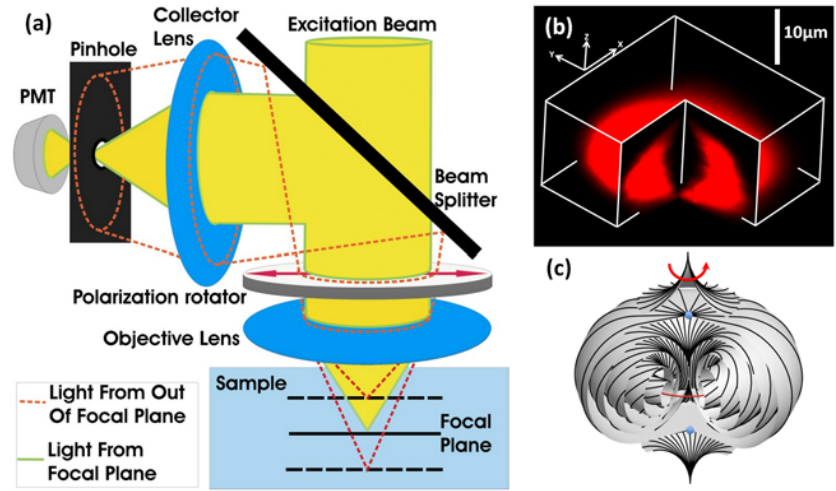

**Figure 10.** Fluorescence confocal polarizing microscopy: (a) a schematic diagram of FCPM with a polarization rotator; (b) 3D image of the T3-1 toron director configuration obtained using FCPM with circularly polarized probing light; (c) a schematic representation of T3-1 configuration with the double-twist cylinder looped on itself and accompanied by two hyperbolic point defects (blue dots).

The basic principle of the FCPM is illustrated with the help of a schematic shown in Fig. 10(a). Let us assume that the transition dipoles of both excitation and fluorescence are parallel to the long axis of the dye molecules. The linearly polarized light incident on the sample causes fluorescence of the dye molecules. The efficiency of light absorption and intensity of detected fluorescence is determined by the angle $\beta$ between the polarization of incident light and the long axis of the dye molecules. Thus, the intensity of the detected fluorescence signal is proportional to $\cos^4\beta$ (in the case of collinear polarized detection) [10]. Fluorescence light emitted from the focal spot passes through the pinhole located in the focal plane of the collector lens that is conjugate to the focal plane of the objective. Signals from out-of-focus regions are prevented from reaching the detector, similar to the case of FCM. To obtain 3D image of the sample, the tightly focused laser beam raster-scans in the lateral plane (perpendicular to the microscope's optical axis), and then, by moving the objective (or the sample-stage) step-wise along the axial direction, the scan is repeated at each axial depth. As a result, the resultant 3D image is comprised of a stack of thin (submicron) horizontal optical slices stored in the computer memory that can be then software-processed and presented in a variety of different formats.

FCPM offers detailed 3D visualization of orientational ordering in soft matter systems like liquid crystals. The example of 3D FCPM texture shown in Fig. 10(b) visualizes the structure of the so-called "toron" [Fig. 10(c)] [14] generated by an infrared laser beam in a cholesteric liquid crystal sandwiched between two glass substrates with vertical boundary conditions. The texture shown in Fig. 10(b) was obtained by using FCPM with circularly polarized probing light. A schematic representation in Fig. 10(c) shows the reconstructed director configuration of the toron with a double-twist cylinder looped on itself and accompanied by two hyperbolic point defects. FCPM imaging can thus visualize the static equilibrium structures of long-range molecular alignment in soft matter systems. On the other hand, the fast FCPM setup, such as the one based on a Nipkow disc confocal microscopy and an achromatic polarization rotator (Fig. 9), can be used to visualize dynamics of 3D director fields in liquid crystals.

### 10.9. Nonlinear optical microscopy

Nonlinear optical (NLO) techniques utilize intrinsic and/or extrinsic nonlinear optical responses of materials such as biological systems and a variety of soft matter systems [15]. These imaging techniques are based on nonlinear light-matter interactions that result in either emission, such as multi-photon excitation fluorescence/luminescence, or scattering, such as second harmonic generation (SHG) [16], sum frequency generation (SFG) [17], third harmonic generation (THG) [18], and coherent Raman scattering, viz., coherent anti-Stokes Raman scattering (CARS) [19,20,21,22] and stimulated Raman scattering (SRS) [23,24].

Nonlinear optical microscopy has recently emerged as a powerful tool for non-invasive, label-free imaging with high 3D resolution capable of probing highly scattering thick biological and soft matter samples. In contrast to conventional single-photon excitation schemes (such as those used in fluorescence microscopy and ordinary Raman microscopy), the NLO signal is generated from nonlinear optical interactions involving multiple excitation photons. Hence, there are several advantages of using NLO microscopy as compared to conventional (linear) optical imaging: (1) low out-of-focus photobleaching (multi-photon absorption occurs only at the focus), (2) low photodamage (for example biological tissues absorb less in the near-infrared), (3) ability to excite ultra-violet (UV)-excitable fluorophores with visible or near-infrared light sources via two- or three-photon absorption, (4) inherent optical sectioning (no pinhole required because of a small excitation volume at the focal spot), (5) ability to work with thick specimens (larger penetration depth) and (6) chemical bond selectivity in coherent Raman imaging. However, nonlinear optical microscopy typically requires more expensive pulsed lasers (typically a femtosecond pulsed laser) and more complicated microscopy setups that are rarely available commercially.

Multi-photon excitation is a nonlinear process typically associated with absorbing two, three, or more photons of near-infrared light and emitting a single photon at a wavelength shorter than the excitation wavelength. The comparison of energy diagrams of the single- and multi-photon processes is shown in Fig. 11(a). Nonlinear processes in an optically excited nonlinear medium can be described by the induced polarization $P(t)$ expressed as a power series in electric field $E(t)$:

$$P(t) \propto \chi^{(1)} E(t) + \chi^{(2)} E^2(t) + \chi^{(3)} E^3(t) + \cdots,$$

where the coefficients $\chi^{(n)}$ are the *n*-th order susceptibilities of the medium [15]. This expression can be used to describe a number of NLO processes discussed below from the standpoint of their use for imaging purposes.

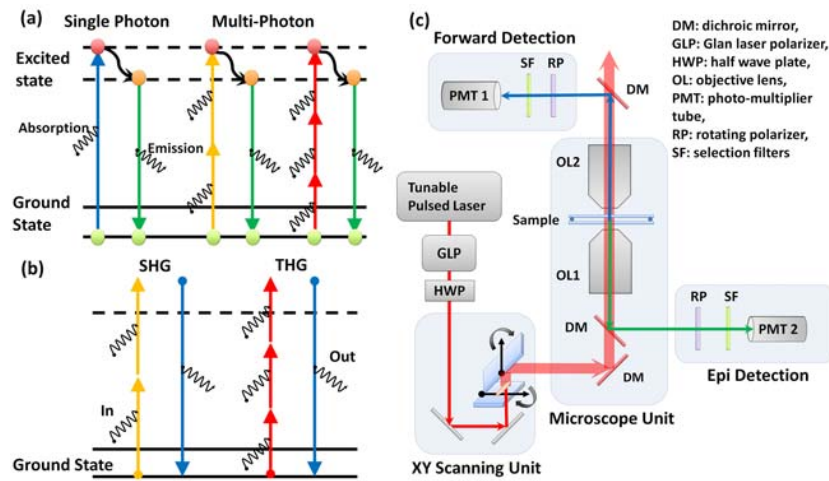

**Figure 11.** Multi-photon excitation fluorescence and multi-harmonic generation microscopy. Energy diagram of (a) single and multi-photon processes and (b) second and third harmonic generation processes. (c) A schematic diagram of nonlinear optical imaging setup based on a tunable pulsed laser, xy scanning mirror, and inverted microscope with multiple detection channels.

In this section, we first describe multi-photon excitation fluorescence microscopy and multi-harmonic generation microscopy and then proceed to the discussion of molecular imaging techniques utilizing coherent Raman scattering, such as recently introduced CARS microscopy, CARS polarizing microscopy, and SRS microscopy.

### 10.9.1. Multi-photon excitation fluorescence microscopy

Compared to the single photon fluorescence technique, multi-photon excitation fluorescence microscopy requires higher peak power of excitation. A femtosecond or picosecond pulsed laser is typically used as an excitation light source with wavelength in the near-infrared region. The intensity of the excitation light falls off inversely as the square of the axial distance from the focal plane and the efficiency of multi-photon absorption away from the focal point is extremely low because the probability of exciting a fluorophore falls off inversely as the fourth power of the axial distance [15]. Unlike in FCM, the detection of fluorescent light does not require a pinhole, because the excitation volume is inherently small enough to enable high resolution imaging in 3D. The use of a near-infrared excitation laser typically helps to image thick samples with deep penetration and less photodamage to the sample.

A simple schematic diagram of multi-photon excitation fluorescence microscopy setup is shown in Fig. 11(c). The setup utilizes a tunable pulsed laser, an $xy$ scanning mirror, an inverted microscope with multiple detection channels, and a rotating polarizer. Two different nonlinear optical imaging modes can be implemented simultaneously, for example, by collecting the fluorescence signal in the backward detection (also known as epi-detection) and the second harmonic generation signal in the forward detection, as shown in Fig. 11(c).

Polarization sensitive excitation and detection of NLO signals is very useful in the study of anisotropic materials such as liquid crystals for imaging of 3D patterns of long-range molecular orientation. An example of multi-photon excitation fluorescence polarizing microscopy in Fig. 12 shows a three-photon excitation fluorescence image of a colloidal particle in a smectic liquid crystal [22] (note that the liquid crystal molecules in this particular case serve as fluorophores themselves). The intensity of detected nonlinear optical polarizing microscopy signals depends on the angle $\beta$ between the polarization of excitation pulses and the liquid crystal director **n(r)** as $\sim \cos^{2m} \beta$ for the detection with no polarizer and as $\sim \cos^{2(m+1)} \beta$ for imaging with the polarizer in the detection channel collinear with the polarization of the excitation beam, where *m* is the order of the nonlinear process (for example *m*=2 for two-photon excitation and *m*=3 for three-photon excitation [20,21,22]).

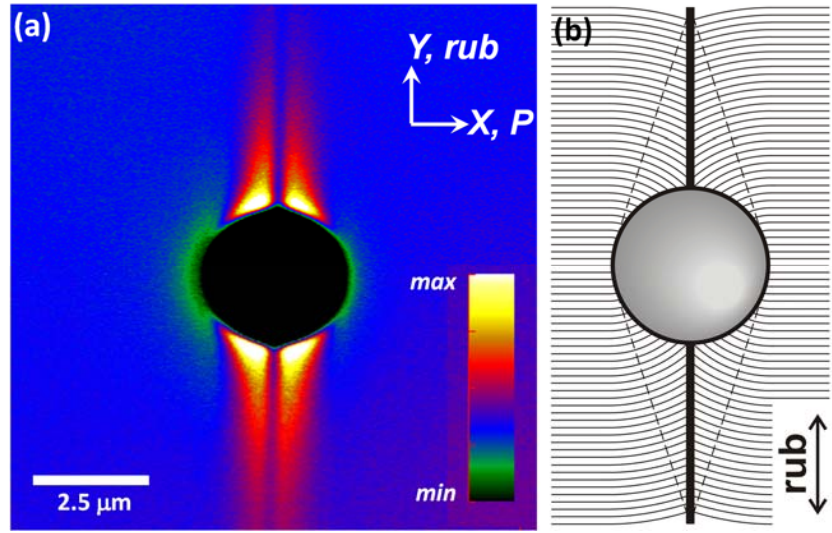

**Figure 12.** Application of multi-photon excitation fluorescence microscopy to imaging of a colloidal particle in a smectic liquid crystal. (a) Three-photon excitation fluorescence image of smectic liquid crystal layers deformations around a melamine resin sphere obtained with excitation at 870 nm and detection within 390~450 nm. (b) Reconstructed deformations of smectic layers around a spherical inclusion.

### 10.9.2. Multi-harmonic generation microscopy

The second- and third-harmonic generation microscopies derive contrast from variations in a specimen's ability to generate the respective harmonic signal from the incident light [16,17,18]. These multi-harmonic generation microscopies also require an intense laser light source (i.e., a pulsed laser). SHG signal emerging from the material is at exactly half the wavelength (frequency doubled) of the incident light interacting with the material via the second order nonlinear process. While two-photon excitation fluorescence is also a second order nonlinear process, it loses some energy during relaxation from the excited state and, therefore, fluorescence occurs at wavelengths longer than half of excitation wavelength. In contrast, SHG process is energy conserving. The energy diagram of second and third harmonic generation processes [Fig. 11(b)] shows that there is no energy loss in the scattered light after multiple photons combine into a single photon of higher energy. In general, the setup for multi-harmonic generation microscopy is similar to that of multi-photon excitation fluorescence microscopy. Therefore, multi-harmonic generation imaging can be implemented with detection in the forward channel and simultaneously multi-photon excitation fluorescence imaging can be collected in the epi-detection channel, as shown in Fig. 11(c). SHG does not depend on excitation of fluorescent molecules. Hence it is not required to tag the specimen with a dye and the effects of photobleaching and photodamage are avoided. The second-order term $\chi^{(2)}$ in nonlinear process is nonzero only in media with no inversion symmetry while the $\chi^{(3)}$ term is nonzero for all media. Thus, imaging can also provide information about the symmetry of the studied materials by using these nonlinear processes.

As an example of the use of SHG microscopy in the study of soft matter, Figure 13 shows SHG images of a smectic C* liquid crystal obtained by using excitation at 1050nm and detection at 525nm for two orthogonal polarizations of excitation light. The spectra and selection filters corresponding to SHG images are shown in Fig. 13(a). A strong SHG signal reveals the polar ordering of the smectic C* phase with focal conic domains [22].

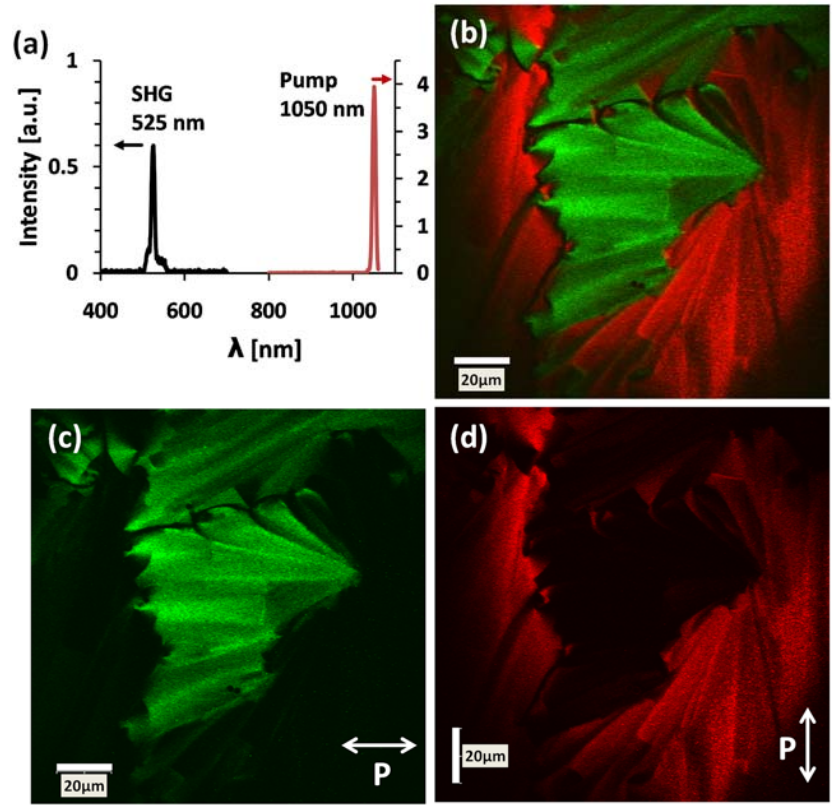

**Figure 13.** Second harmonic generation imaging of a smectic C* liquid crystal. (a) A spectrum showing the excitation pulse and the generated SHG signal. (b) In-plane colocalized superimposed texture of two SHG images (c,d) that were obtained separately for two orthogonal polarizations.

### 10.9.3. Coherent anti-Stokes Raman scattering microscopy

Coherent anti-Stokes Raman scattering microscopy is a non-invasive, label-free nonlinear imaging technique that utilizes molecular vibrations to obtain imaging contrast [19,20,21,22]. CARS microscopy is often compared to conventional Raman microscopy, as both techniques probe the same Raman active modes. Spontaneous Raman scattering can be induced by a single continuous wave laser, whereas CARS requires at least two pulsed laser sources at different frequencies. The spontaneous Raman signal is typically detected on the red side of the spectrum compared to the excitation radiation, where it might be difficult to discriminate it from fluorescence signal. The CARS signal is detected on the blue side, which is free from fluorescence interference, but it typically comes with a non-resonant background contribution.

The CARS technique utilizes a third order nonlinear process which involves three photons at two different frequencies called pump/probe ($\omega_p = \omega_{pump} = \omega_{probe}$) and Stokes ($\omega_s = \omega_{stokes}$) beams as shown in Fig. 14(a). When the frequency difference between the pump/probe and the Stokes beams matches a certain molecular vibrational frequency ($\omega_{vib} = \omega_p - \omega_s$) and the phase-matching condition of three input photons is fulfilled, a strongly enhanced blue-shifted anti-Stokes ($\omega_{as} = 2\omega_p - \omega_s$) resonance signal is generated in the sample. The resultant CARS signal is at a wavelength shorter than that of the pump/probe and Stokes beams, as shown in Fig. 14(b). The CARS intensity scales with the intensities of the excitation beams as follows

$$I_{CARS}\left(2\omega_p - \omega_s\right) \propto \left(\chi_{CARS}^{(3)}\right)^2 I_p^2(\omega_p) I_s(\omega_s),$$

where $\chi_{CARS}^{(3)}$ is the third order susceptibility, $I_p$ and $I_s$ are the intensities of the pump/probe and Stokes beams, respectively.

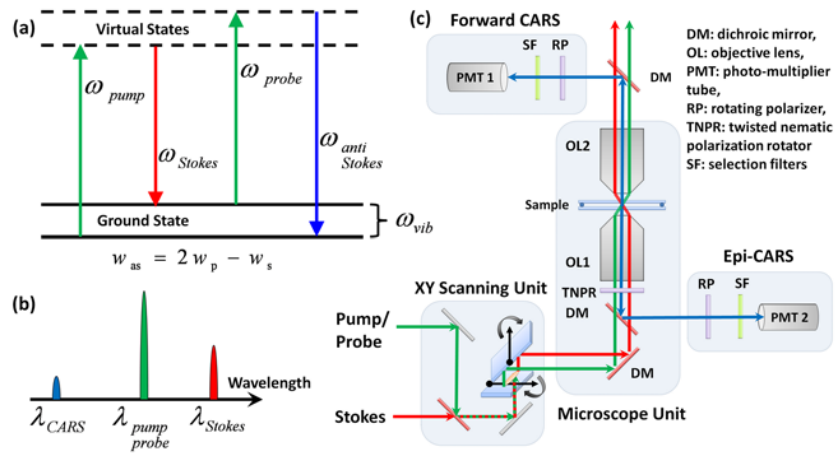

**Figure 14.** Coherent anti-Stokes Raman scattering microscopy. (a) The energy diagrams of CARS at $\omega_{as}=2\omega_p$-$\omega_s$ when $\omega_{vib}=\omega_p$-$\omega_s$. (b) CARS signal generated at shorter wavelength than the pump and Stokes wavelengths. (c) A schematic diagram of CARS-polarizing microscopy setup utilizing the synchronized pump/probe and Stokes pulses, xy scanning galvano-mirrors, and inverted microscope with forward and epi-detection channels.

A simple schematic diagram of CARS microscopy setup based on synchronized pump/probe and Stokes beams, $xy$ scanning mirror, and an inverted microscope with forward and epi-detection channels is shown in Fig. 14(c) [22]. There are several options for a light source in CARS microscopy: (1) two tightly synchronized pulsed lasers, (2) a synchronously-pumped intra-cavity doubled optical parametric oscillator, (3) a single femtosecond laser pulse spectrally shaped to select two frequencies $\omega_p$ and $\omega_s$, or (4) a synchronously-generated supercontinuum (by using highly nonlinear fiber) with filter-selected $\omega_p$ and/or $\omega_s$ from a single femtosecond laser. CARS microscopy has several unique benefits in the study of soft matter and biological samples: (1) intrinsic vibrational contrast (no labeling needed), (2) a strong, directional signal (CARS is more sensitive than conventional vibrational microscopy), and (3) 3D sectioning capability with less photodamage, no photobleaching, and deeper penetration in thick turbid media, similar to other nonlinear optical microscopy techniques.

### 10.9.4. Coherent anti-Stokes Raman scattering polarizing microscopy

As discussed earlier, polarization sensitive imaging is an essential tool for imaging soft matter systems possessing long-range orientational order. CARS polarizing microscopy setup [Fig. 14(c)] has a polarization control based on a twisted nematic achromatic polarization rotator for excitation and epi-detection of CARS signals and an optional rotating polarizer in the forward-detection channel. Since CARS is a three-photon process, the dependence of the CARS signal on the angle $\beta$ between the polarization of excitation light and the director of sample is $\propto \cos^6 \beta$ with no polarizer in the detection channel and $\propto \cos^8 \beta$ with a polarizer collinear to excitation polarization in the detection [20,21,22].

Similar to other NLO polarizing microscopy techniques, CARS polarizing microscopy exhibits a stronger sensitivity to spatial variations of the director **n(r)** as compared with the single-photon FCPM imaging.

As an example of the use of CARS polarizing microscopy technique, Figure 15(a,b) shows a vertical cross-section and in-plane images of a toric focal conic domain in a smectic liquid crystal thin film on a solid substrate. The signal due to CN-triple-bond vibration of the smectic liquid crystal (4-cyano-4'-octylbiphenyl, 8CB) provides information about the spatial pattern of the liquid crystal director. When the director is parallel to the polarization of excitation light, the intensity of the CARS signal is maximized, whereas the CARS intensity is minimized when the director is perpendicular to the polarization of incident light. The reconstructed structure of smectic layers in the toric focal conic domain is shown in Fig. 15(c), which was obtained using the 3D CARS images [Fig. 15(a,b)].

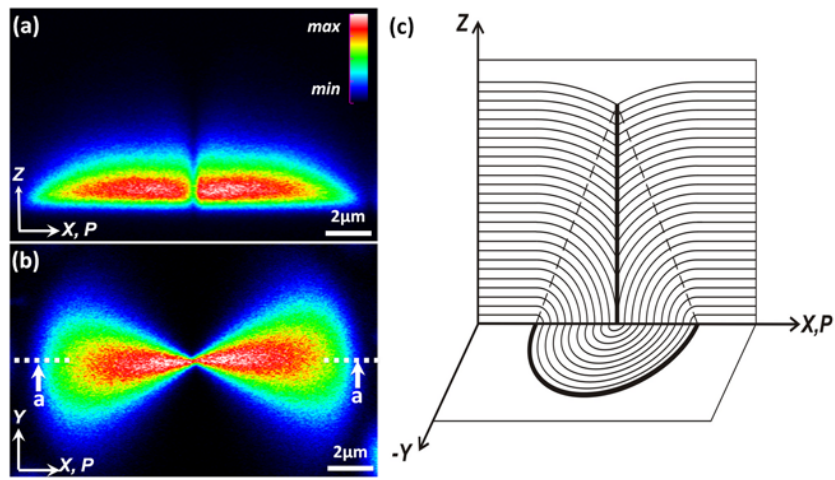

**Figure 15.** Coherent anti-Stokes Raman scattering images of toric focal conic domain in a smcetic liquid crystal thin film on a solid substrate. (a) Vertical cross-section and (b) in-plane images. (c) Reconstructed structure of smectic layers in the toric focal conic domain.

## 10.9.5. Stimulated Raman scattering microscopy

Stimulated Raman scattering is a technique in which two beams, pump (at frequency $\omega_p$) and Stokes (at frequency $\omega_s < \omega_p$), combine to amplify the Stokes Raman signal when the difference between $\omega_p$ and $\omega_s$ equals the vibrational frequency of a certain chemical bond in the molecules comprising the material as shown in Fig. 16(a). Since there are two photons involved, SRS is a second order nonlinear optical process and the intensity of the SRS signal is proportional to the product of the intensities of the pump and the Stokes beams. As a result of the amplification of the Raman signal, the pump beam experiences a loss in its intensity (stimulated Raman loss) while the Stokes beam experiences a gain in its intensity (stimulated Raman gain). Since the gain/loss mechanism occurs only when the difference between the frequencies of the pump and Stokes beam equals the molecular vibration frequency, the non-resonant background is significantly reduced compared to that of CARS.

The relative Raman gain/loss is a very small fraction of the excitation signal intensities. Therefore, a typical implementation of SRS microscopy involves modulation of one of the excitation beams (e.g., Stokes), as shown in Fig. 16(b). The pump beam experiences Raman loss only when the Stokes pulse is present; thus the Raman loss signal in the pump occurs at the Stokes modulation rate, which is then selectively detected by a photodiode with the help of a lock-in amplifier. A simplified representative set-up is shown in Fig. 16(c). Similar to CARS, SRS can also be implemented to image multiple molecular functionalities simultaneously. This is demonstrated in the example in Fig. 16(d), where a drug penetration enhancer dimethyl sulfoxide (DMSO) is imaged (shown in green) together with lipid (shown in red), demonstrating that DMSO is insoluble in the lipid [23,24]. Similar to CARS and other NLO imaging techniques, SRS can be extended to enable orientation sensitive imaging by means of polarized excitation and detection.

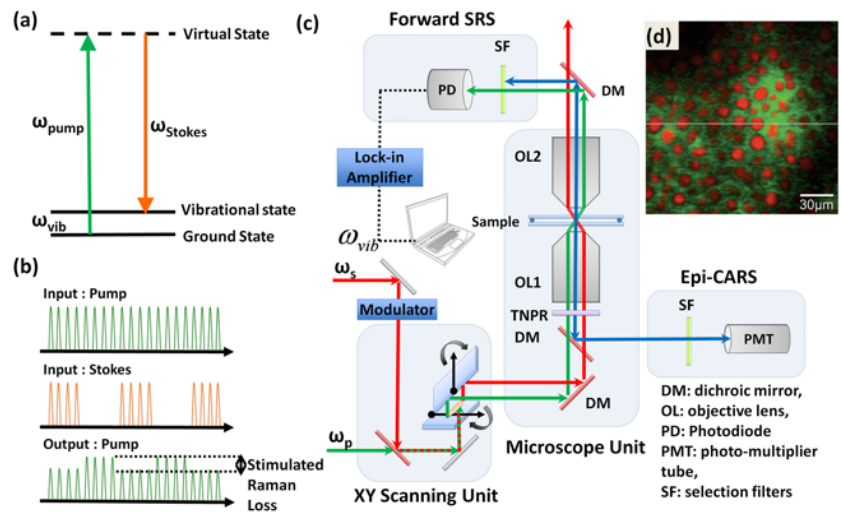

**Figure 16.** Stimulated Raman scattering microscopy. (a) Energy diagram and (b) detection scheme for SRS. Stokes beam is modulated at high frequency and the resulting amplitude modulation of the pump pulse (stimulated Raman loss) can be detected. (c) A schematic diagram for SRS microscope with forward SRS and epi CARS detection. (d) Two-color SRS image of DMSO (green) and lipid (red) in the subcutaneous fat layer. (image reprinted from [23] with permission from AAAS).

## 10.10. Three-dimensional localization using engineered point spread functions

Along with imaging of materials, high resolution tracking and localization of molecules and particles is helpful for understanding their self-assembly in soft matter systems. For instance, the image blurring caused by displacing the object away from the focal plane can be used as a rough measure in determining the axial position of an object in an image. One approach to localize an object in 3D with much finer resolution is to design the point spread function of the imaging system, as opposed to relying on the standard Airy disc PSF (Fig. 1c). Figure 17(a) shows an example of implementation of such an imaging system where the PSF is made up of two lobes that rotate as they traverse along the microscope's optical axis, forming a "double-helix" near the focal point [Fig. 17(b)]. The setup is built around a phase-mask at the Fourier plane in the imaging path, which can be programmed by use of a spatial light modulator (SLM), as shown in Fig. 17(c) [25]. The SLM modulates the phase of the imaging beam at its Fourier plane, so that the resultant image of a particle captured by the CCD camera is presented in the form of two lobes. The axial position of an imaged particle is encoded as the angle of the line joining the lobes [Fig. 17(d)]. This technique can be used to localize scatterers (e.g., colloids) or fluorescent emitters (e.g., single fluorescent molecules) with the help of appropriate dichroic mirror and filters [Fig. 17(e)] achieving nanoscale resolution (typically ~10nm) in this spatial localization technique [26].

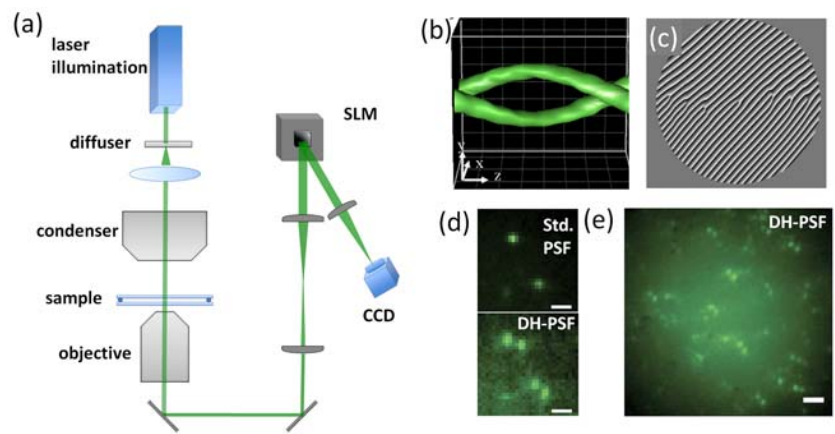

**Figure 17.** Optical imaging by use of engineered PSF. (a) Optical set-up for imaging with pre-engineered PSF. (b) Schematic representation of double-helix PSF. (c) Typical phase mask used to generate double helix-PSF. (d-e) Single molecule detection in 3D using double helix-PSF technique. (images from [25] copyright 2009 National Academy of Sciences, U.S.A.)

## 10.11. Integrating three-dimensional imaging systems with optical tweezers

Optical manipulation has proved to be of great importance in the study of colloidal systems, liquid crystals, and biological materials. Optical tweezers have been employed to study various liquid crystal systems, colloidal assemblies and interactions, as well as topology and structure of defects [27]. Manipulation of foreign inclusions and defects gives rise to a variety of director patterns in liquid crystals, which requires 3D imaging to fully understand their structures [14].

An integrated optical imaging and manipulation system, such as the one schematically depicted in Fig. 18(a), allows for simultaneous non-contact optical manipulation and imaging [28]. The optical manipulation employs a liquid crystal based SLM, which spatially modulates the phase of the incident trapping laser beam. The phase modulation is controlled in real time by programming the SLM to display holograms corresponding to the specified trap pattern at video rate. The lenses in the optical train form a 4f telescope, so as to image the plane of the SLM at the back-aperture of the trapping objective. The main advantage of SLM based holographic optical tweezers is that it allows simultaneous optical manipulation of multiple particles in the lateral as well as the axial directions. The SLM also allows generation of non-Gaussian beams (e.g., Laguerre-Gaussian beams), which can be used to generate liquid crystal director patterns under the effect of various unconventional light-intensity profiles [14].

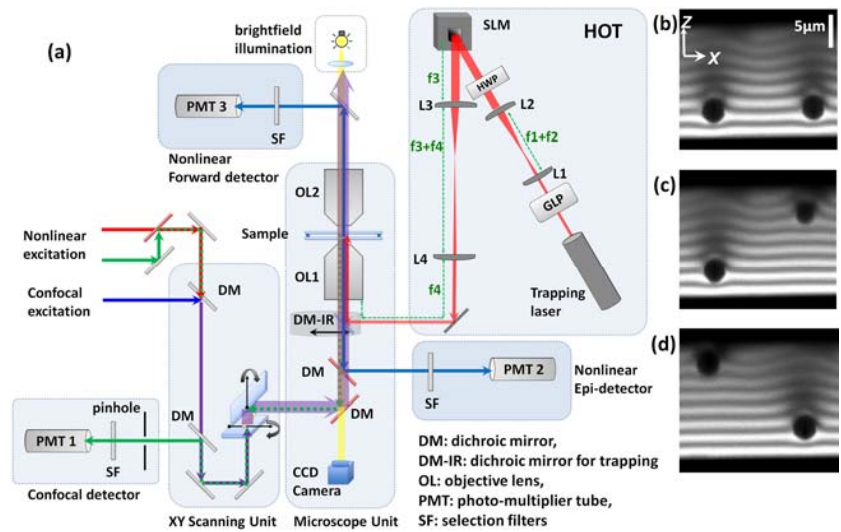

**Figure 18.** Integrated setup of 3D imaging and holographic optical trapping. (a) A schematic diagram of 3D optical manipulation with holographic optical trapping (HOT) and simultaneous imaging with FCPM and nonlinear optical microscopy. (b, c, d) FCPM vertical cross-sections showing spherical particles in cholesteric liquid crystal moved axially and imaged using the integrated setup.

Simultaneous 3D optical manipulation and imaging capabilities of the system are demonstrated in Fig. 18(b-d), where colloidal particles immersed in a cholesteric liquid crystal are moved axially perpendicular to the cholesteric layers, and the layer configurations are imaged in vertical cross-sections with FCPM [28]. The cholesteric layer deformations caused by the presence of particles can be clearly seen in the FCPM images. The integration of 3D non-contact manipulation with 3D imaging enables the design and characterization of soft-matter composites comprising colloids, nanoparticles, liquid crystals, polymers, etc. The integrated setup allows one to uncover the nature of interactions between various components of the materials system as well as control of their assembly and organization. In addition, integration with an engineered PSF tracking and localization technique can expand functionality of the system to probe molecular and colloidal interactions with higher precision.

### 10.12. Outlook and perspectives

Some of the grand challenges from the standpoint of the use of optical microscopy in the study of soft matter systems include the need of nanoscale spatial resolution (to directly probe molecular and colloidal self-assembly in material systems), sub-millisecond temporal resolution (to probe fast dynamic processes), sensitivity to different chemical bonds of molecules comprising soft matter systems, and the ability to visualize orientations of molecules and chemical bonds comprising them. Out of all of these challenges, achieving nanoscale resolution of optical imaging is, perhaps, the most important. However, in terms of resolution, optical imaging has undergone only incremental improvements over centuries of its use, even though imaging techniques are most frequently used not only in soft matter research but also in many other branches of science. Novel imaging approaches with improved resolution are especially needed to further the collective understanding of soft matter systems on the nanoscale. Several recently introduced approaches show the promise of providing significant breakthroughs in terms of overcoming resolution limits and may also be applied to the study of soft matter systems. In the future, their further development may allow for direct imaging of the self-assembly of molecules and nanoparticles and their condensed phase behavior.

Recently, a bulk of theoretical and experimental research has shown the feasibility of diffraction-unlimited optical imaging by use of simple optical elements based on metamaterials with a negative index of refraction. Theoretical works by Veselago [29] and Pendry [30], followed by numerous theoretical and experimental explorations, have demonstrated that the diffraction limit of conventional lenses is not a limiting factor for focusing of light by a lens made of a flat slab of metamaterial with a negative index of refraction, often referred to as a "super-lens" or a "perfect lens." The improved resolution of the negative-index super-lens is due to the transmission of the evanescent surface waves, which are not lost (unlike in the case of conventional lenses). Being first considered theoretically over four decades ago by Veselago [29], metamaterials (also called "left-handed materials") have never been encountered in nature or fabricated until the theoretical work of Pendry [30] provided important physical insights into how these unusual materials can be realized. Metamaterials are composed of structural units much smaller than the wavelength of incident light, so they appear homogeneous to the electro-magnetic waves. To obtain metamaterials with negative refractive index at optical frequencies, one has to achieve simultaneous "negative" electric and magnetic response in the artificial nanofabricated composite material. Although the practical uses of metamaterials are still hindered by a number of technical challenges, such as losses, lenses and other optical elements made of metamaterials may one day revolutionize optical microscopy and enable nanoscale diffraction-unlimited imaging [31].

Several new techniques enabling sub-diffraction limited high-resolution optical imaging include near-field scanning optical microscopy (NSOM) [32], photoactivated localization microscopy (PALM) [33,34], and stochastic optical reconstruction microscopy (STORM) [35]. They can be extended for the study of soft matter systems, including the orientation-sensitive imaging of long-range molecular alignment patterns in liquid crystals. NSOM is a type of scanning probe microscopy for nanoscale investigation by observing the properties of evanescent waves. This is done by placing the probe very close to the sample surface. Light passes through a sub-wavelength diameter aperture and illuminates a sample that is placed within its near field, at a distance smaller than the wavelength of the light. With this technique, the resolution of the image is limited by the size of the detector aperture and not by the wavelength of the illuminating light. In particular, lateral resolution of 20 nm and vertical resolution of 2–5 nm have been demonstrated [32]. However NSOM is difficult to operate in a non-invasive mode and has a limited imaging depth.

The basic concept of PALM and STORM techniques is to fill the imaging area with many non-fluorescing fluorophores that can be photoactivated into a fluorescing state by a flash of light [33,34,35]. Since this photoactivation is stochastic, only a few, well-separated molecules can be detected and then localized by fitting the PSF with Gaussians with high precision. This process is repeated many times while photoactivating different sets of molecules and building up an image molecule-by-molecule. The resolution of the final reassembled image can be much higher than that limited by diffraction. The demonstrated imaging resolution is 20 to 30 nm in the lateral dimensions and 50 to 60 nm in the axial dimension [36]. However the major problem with these emerging techniques is that it takes time on the order of hours to collect the data to get these high resolution images. Study of soft matter systems such as biological structures and colloidal nanoparticle dispersions will tremendously benefit from further development and use of these techniques.

In conjunction with high spatial resolution, imaging systems with high temporal resolution are required to study dynamic processes in soft-matter systems, for instance, the time-evolution of self-assembled structures, effect of applied external fields, etc. Nowadays, high-speed cameras with capture rates of up to a million frames per second are commercially available. There has been significant development on the front of enabling nonlinear, non-invasive optical imaging modalities like CARS and SRS to operate at video-rate, which will immensely benefit biological and soft matter systems research. This will eventually allow for in vivo optical imaging with molecular selectivity [37] and direct characterization of self-assemblies in composite systems.

A variety of optical microscopy techniques discussed here have driven the research in the field of soft matter systems to new frontiers. The older microscopy techniques have allowed 2D imaging and hence the actual 3D configuration of material components composing soft-matter systems was often left to an educated guess [6]. The introduction of 3D imaging techniques has facilitated unambiguous determination of the molecular and colloidal arrangement with precise localizing of other elements. The advent of pulsed lasers has enabled implementation of nonlinear optical microscopies, furthering this trend and rendering the use of labeling agents unnecessary, while at the same time giving similar or better spatial resolution. NLO microscopy has also made it more convenient to image composite soft-matter systems by allowing imaging of different constituents in different nonlinear modalities simultaneously, instead of going through the difficult, and often impossible process of finding appropriate dyes that would bind to each of the constituents selectively. Further general development of optical microscopy techniques is an ongoing quest that will continue to contribute to the body of knowledge of soft matter systems.